# Aggregate dust particles at comet 67P/Churyumov–Gerasimenko

Mark S. Bentley[1*], Roland Schmied[1*], Thurid Mannel[1,2*], Klaus Torkar[1], Harald Jeszenszky[1], Jens Romstedt[4], Anny-Chantal Levasseur-Regourd[3], Iris Weber[5], Elmar K. Jessberger[5], Pascale Ehrenfreund[6,7], Christian Koeberl[8,9], Ove Havnes[10]

1) Space Research Institute, Austrian Academy of Sciences, Schmiedlstrasse 6, 8042 Graz, Austria

2) Physics Institute, University of Graz, Univeristätsplatz 5, 8010 Graz, Austria

3) LATMOS- IPSL, UPMC (Sorbonne Universités), 4 place Jussieu, 75005 Paris, France

4) European Space Research and Technology Centre, Future Missions Office (SREF), Noordwijk, Netherlands

5) Institut für Planetologie, Universität Münster, Wilhelm-Klemm-Str. 10, 48149 Münster, Germany

6) Leiden Observatory, Postbus 9513, 2300 RA Leiden, The Netherlands

7) Space Policy Institute, George Washington University, 20052 Washington, DC, USA

8) Department of Lithospheric Research, University of Vienna, Althanstrasse 14, 1090 Vienna, Austria

9) Natural History Museum, Burgring 7, 1010 Vienna, Austria

10) Department of Physics and Technology, UiT The Arctic University of Norway, N-9037 Tromsø, Norway

* These authors contributed equally to this work

**Comets are believed to preserve almost pristine dust particles, thus providing a unique sample of the properties of the early solar nebula. The microscopic properties of this dust play a key role in particle aggregation during Solar System formation[1,2]. Prior to Rosetta cometary dust was considered to comprise irregular, fluffy agglomerates based on interpretation of remote observations in the visible and infrared[3–6] and study of chondritic porous interplanetary dust particles (IDPs)[7], believed but not proven to originate in comets. Although the dust returned by the Stardust mission[8] has provided detailed mineralogy of particles from comet 81P/Wild, the fine-grained aggregate component was strongly modified during collection[9]. Rosetta offered a unique opportunity to determine, for the first time, the micro-structure of cometary particles *in situ*. Here we show that cometary dust particles are aggregates of smaller, elongated, grains even at the sub-micrometre scale, with structures at distinct sizes indicating hierarchical aggregation. Topographic images of selected dust particles from one to a few tens of micrometres in size show a variety of morphologies, ranging from compact single grains to large porous aggregate particles, similar to chondritic porous IDPs. These observations not only validate the aggregate model of cometary dust, they are also an important input for understanding comets and their formation. The measured grain elongations are similar to the value inferred for interstellar dust and support the idea that such grains could represent a fraction of the building blocks of comets. In the subsequent growth phase, hierarchical agglomeration can be a dominant process[10] and would produce aggregates that stick more easily at higher masses and velocities than homogenous dust particles[11]. The presence of hierarchical dust aggregates in the surface layers of the nucleus also provides a mechanisms for lowering tensile strength and aiding dust release[12].**

MIDAS, the Micro-Imaging Dust Analysis System[13,14], is the first spaceborne atomic force microscope (AFM) and a unique instrument designed to measure the size, shape, texture and microstructure of cometary dust. Flying on the Rosetta spacecraft, it collects dust on sticky targets during passive exposures and images its 3D topography with an unprecedented nanometre to micrometre resolution[13].

Cometary dust was first collected in mid-November 2014. In this work we focus on particles collected from then until the end of February 2015. The collected particles cover a range of sizes from tens of micrometres down to a few 100 nanometres, and have various morphologies, from single grains to aggregate particles with different packing densities. Five examples are presented here.

Figure 1 shows topographic images (height fields) of three particles (A, B and C). Particles A and C will be referred to as compact, since their sub-units (hereafter grains) are tightly packed, and B appears to be a homogeneous grain. The next example (D) is also a compact particle scanned with a higher lateral resolution of 80 nm (Fig. 2) - a factor four better than the previous scan. The final particle (E), presented in Fig. 3, is best described as a loosely packed "fluffy" aggregate comprising many grains. Detailed collection times and geometries for all particles can be found in Extended Data Fig. 1-3.

Aided by the 3D nature of the data, individual grains can be identified, as shown in Fig. 1b, Fig. 2b and Fig. 3b. The properties of these particles and their grains are summarised in Tab. 1 for particles A-D, and in Fig. 3d for particle E. Since particle E extends beyond the edge of the scanned area only lower limits for its dimensions can be given. All further calculations and discussion refer only to this visible region.

Compact particles A and C are both approximately 5.6 µm in effective diameter (hereafter size - see Methods), and are built from grains in the size range $1.93^{+0.10}_{-1.22}$ µm to $3.31^{+0.06}_{-1.23}$ µm. The compact grain B is $2.76^{+0.07}_{-0.61}$ µm in size, comparable to the dust grains of particles A and C. In fact, the topographic image suggests that this grain was originally part of C but detached on impact with the target. Particle D is $1.09^{+0.01}_{-0.25}$ µm in size, again similar to the grains in A-C. However, the higher resolution reveals that this micrometre-sized particle is itself an aggregate of smaller units; seven grains can be resolved, with sizes ranging from $260^{+50}_{-120}$ nm to $540^{+20}_{-250}$ nm. The visible part of particle E has a maximum extent of 14 µm in X and 37 µm in the Y direction. Analysis of its component grains (Fig. 3d) shows sizes in the range from $0.58^{+0.15}_{-0.20}$ to $2.57^{+004}_{-0.51}$ µm with the grain heights ranging between 0.2 µm and 3 µm with 90% smaller than 1.7 µm. These measurements are the first evidence for a continuation of the aggregate nature of dust particles below the size range observed by COSIMA (10s-100s micrometres)[15].

Particle E also shows a morphology strongly reminiscent of stratospheric chondritic porous (CP) IDPs, long suspected of having a cometary origin. This link is consistent with observations by COSIMA for larger dust particles, which also measured similar compositions for dust at 67P and IDPs[15,16]. One notable difference to

IDPs is the extremely flat nature of particle E, which has a height an order of magnitude lower than its (minimal) lateral dimension. Indeed all of the particles presented here have flattened shapes to some degree (see Tab. 1). It is not yet clear if this is an intrinsic property of cometary dust or the result of a rearrangement of grains on impact. COSIMA has observed that sub-mm aggregate particles undergo rearrangement of their grains on impact, producing flattened shapes[15]. Additionally, COSIMA collected small, apparently compact particles which are also flat, but the resolution is insufficient to determine if they are single grains, or indeed aggregates themselves. On the other hand, cluster-cluster aggregation with rotating grains can form elongated structures with very high aspect ratios[17], and laboratory experiments have produced dust "flakes"[18].

Investigation of the size distribution of CP IDPs and fine-grained material returned by the Stardust mission[19,20] showed that the majority of their component grains are smaller than 500 nm[20,21]. Fig. 3d shows that 90% of the grains in particle E are smaller than 2 μm, comparable to particle D. which itself is built from grains smaller than about 500 nm. This suggests that the grains of the fluffy aggregate particle E are also aggregates of sub-micrometre components similar to those in CP IDPs and points towards a hierarchical structure. Hierarchical growth (i.e. aggregates of smaller aggregates) has been proposed as a growth mechanism in the protoplanetary disc when fragmentation of larger particles provides a population of smaller aggregates available for agglomeration[10]. The sticking probability of such particles can be higher for a given mass and velocity and need to be accounted for in dust particle growth models[11]. Hierarchical aggregates have also been invoked to produce a cometary dust layer with sufficiently low tensile strength to allow for dust release[12].

Since MIDAS provides, for the first time, real measurements of the grain shapes, it is possible to go further and evaluate which models support the observations. The elongation of the grains is found by calculating the ratio of the longest and shortest perpendicular axis. Further details are described in the Supplementary Information. For particle E the grain heights are almost all smaller than their in-plane diameters, suggesting that it comprises a single layer of grains, allowing accurate grain heights to be determined. The elongation is calculated for 114 grains (the 11 omitted grains show strong distortions due to tip convolution), giving an average elongation of $2.87^{+1.90}_{-0.44}$ (i.e. the largest axis is 3 times longer than the smallest). The compact particles show similar values (Tab. 1).

Elongated grains are considered in several models of cometary dust. For example Greenberg and Gutasfson[22] suggested that comets aggregate from interstellar grains. They modelled the dust as cylinders with aspect ratios of 2-4 and found good agreement between light scattering experiments and observations. Other authors have similarly found good agreement between simulations using aggregates of spheroidal particles and observational data[23,24]. The elongated nature of interstellar dust can be inferred from linear polarisation of starlight due to partially aligned grains[25]. The core-mantle structure proposed for interstellar and cometary dust[26] cannot be confirmed by MIDAS data alone, but the elongation measurement supports the idea of a common precursor grain, or growth mechanism.

**Supplementary Information** is linked to the online version of the paper at www.nature.com/nature

## Acknowledgements

Rosetta is an ESA mission with contributions from its member states and NASA. We also thank the Rosetta Science Ground Segment and Mission Operations Centre for their support in acquiring the presented data. MIDAS became possible through generous support from funding agencies including European Space Agency, PRODEX programme, the Austrian Space Agency, the Austrian Academy of Sciences and the German funding agency DARA (later DLR). ACLR acknowledges support from the French Space Agency, CNES. TM gratefully acknowledges the Steiermärkische Sparkasse and the Karl-Franzens Universität Graz for their financial support. PE acknowledges support from the NASA Astrobiology Institute. RS thanks Ferdinand Hofer and Harald Plank for their discussions and the Austrian Research Promotion Agency (FFG) for financial support. All data presented in this paper will be made available in the ESA Planetary Science Archive.

## Author contributions

RS, TM and MSB planned the experiments on MIDAS, analysed and interpreted the data and wrote the manuscript. MSB developed the planning and data processing pipelines. RS and TM implemented the elongation calculations. RS performed the post-processing and calibration as well as the particle/grain measurement and is responsible for the graphical data presentation. TM considered the uncertainties for all data. ACLR provided information on cometary dust derived from polarimetric observations and its interpretation. HJ supported the experiments with software updates. All authors discussed the results and commented on the manuscript.

## Author information

Reprints and permissions information is available at www.nature.com/reprints. The authors declare no competing financial interests. Readers are welcome to comment on the online version of the paper. Correspondence and requests for materials should be addressed to MSB (mark.bentley@oeaw.ac.at).

# Tables

|            | type             | d ± Δd [μm]            | $z_{max}$ [μm] | elongation              |
|------------|------------------|------------------------|----------------|-------------------------|
| **particle A** | compact particle | $5.48^{+0.04}_{-1.10}$ | 1.79           | $3.32^{+0.14}_{-0.41}$  |
| **grain 1**    | dust grain       | $3.31^{+0.06}_{-1.23}$ | 1.79           | $2.94^{+0.12}_{-0.43}$  |
| **grain 2**    | dust grain       | $2.62^{+0.08}_{-0.87}$ | 1.33           | $3.04^{+0.15}_{-0.42}$  |
| **grain 3**    | dust grain       | $1.93^{+0.10}_{-1.22}$ | 1.57           |                         |
| **grain 4**    | dust grain       | $2.62^{+0.08}_{-1.07}$ | 1.55           | $1.96^{+0.09}_{-0.40}$  |
| **particle B** | dust grain       | $2.76^{+0.07}_{-0.61}$ | 1.02           | $3.14^{+0.18}_{-0.42}$  |
| **particle C** | compact particle | $5.79^{+0.04}_{-0.87}$ | 1.39           | $4.77^{+0.24}_{-0.50}$  |
| **grain 1**    | dust grain       | $2.66^{+0.07}_{-0.92}$ | 1.33           | $2.26^{+0.11}_{-0.42}$  |
| **grain 2**    | dust grain       | $2.57^{+0.08}_{-0.72}$ | 1.14           | $2.80^{+0.15}_{-0.41}$  |
| **grain 3**    | dust grain       | $2.18^{+0.09}_{-0.83}$ | 1.28           | $2.23^{+0.12}_{-0.39}$  |
| **grain 4**    | dust grain       | $2.42^{+0.08}_{-0.90}$ | 1.39           | $2.31^{+0.11}_{-0.39}$  |
| **grain 5**    | dust grain       | $2.31^{+0.08}_{-0.87}$ | 1.38           | $2.32^{+0.11}_{-0.39}$  |
| **particle D** | compact particle | $1.09^{+0.01}_{-0.25}$ | 0.42           | $3.36^{+0.23}_{-0.47}$  |
| **grain 1**    | dust grain       | $0.26^{+0.05}_{-0.12}$ | 0.17           | $1.89^{+0.19}_{-0.36}$  |
| **grain 2**    | dust grain       | $0.48^{+0.03}_{-0.16}$ | 0.22           | $2.52^{+0.20}_{-0.47}$  |
| **grain 3**    | dust grain       | $0.41^{+0.03}_{-0.14}$ | 0.31           | $1.62^{+0.11}_{-0.27}$  |
| **grain 4**    | dust grain       | $0.33^{+0.04}_{-0.13}$ | 0.25           | $1.74^{+2.51}_{-0.71}$  |
| **grain 5**    | dust grain       | $0.46^{+0.03}_{-0.17}$ | 0.37           | $1.53^{+0.09}_{-0.28}$  |
| **grain 6**    | dust grain       | $0.54^{+0.02}_{-0.25}$ | 0.42           | $2.00^{+5.07}_{-0.82}$  |
| **grain 7**    | dust grain       | $0.26^{+0.05}_{-0.15}$ | 0.32           | $2.00^{+8.03}_{-0.97}$  |

**Table 1: Size, height and elongation data of the presented dust particles (A-D) and their component dust grains .** $d$ is the diameter of a circle with equivalent area, $z_{max}$ is the maximum height above the substrate surface. For particle A grain 3, particle D grain 4, 6 and 7 the maximal elongation is found for the ratio of the two lateral dimensions that are attached with especially large uncertainties. Thus for particle A grain 3 no accurate elongation can be given and for the grains of particle D the elongation is attached with extensive uncertainties.

# Figure legends

**Figure 1**: **AFM topographic images of particles A, B and C and their sub-units.** (a) 20x50 µm overview image with a pixel resolution of 312 nm and a colour scale representing the height. (b) Particle B and the sub-units of particles A and C are outlined in cyan. (c) and (d) 10x10 µm 3D (rotated) images of particles A and C with two times height exaggeration to aid visualisation.

**Figure 2: AFM topographic images of particle D and its sub-units.** (a) 5x5 µm overview image with a pixel resolution of 80 nm and a colour scale representing the height. (b) Outline of the sub-units of particle D as a cyan overlay. (c) 3D (rotated) image of the particle with two times height exaggeration to aid visualisation.

**Figure 3: AFM topographic images of particle E showing its sub-units and their size distribution.** (a) 14x37 µm overview image with a pixel resolution of 210 nm and a colour scale representing height. (b) identified grains have been marked with a cyan outline. (c) 3D 14x34 µm view (rotated and cropped) to aid visualisation (corresponding to the red frame in (a)). (d) Cumulative distribution of the grain equivalent diameters with error bars given in grey. The left scale gives the absolute grain numbers and the right scale is giving the probability for particles to be below the specific values.

## Methods

### Data acquisition and calibration

Exposure durations and times were planned by estimating the dust flux using the predicted spacecraft position, pointing and a dust flux model for 67P derived from observational data[27]. For a graphical visualization of the exposure geometries, see Extended Data Fig. 1-3.

MIDAS operates in a slightly different way than most terrestrial AFMs, by making a careful approach to the sample at each pixel position and then moving away by a so-called retraction distance before moving to the next pixel, resulting in long scan times and possible distortion[13,14]. Distortion correction is performed using scans of on-board calibration targets, and polynomial background correction is used to remove height drifts. This procedure was performed with the data used to produce Fig. 1 and 3. The scan shown in Fig. 2 was much shorter and no significant distortion was observed hence only background subtraction was performed. Particle and grain heights are measured relative to the substrate surface, which is very clear for Fig. 1 and Fig. 2, but the zero reference level had to be set manually for each grain in Fig. 3, since the steps would otherwise distort the measurements.

The lateral extent of both particles and grains is characterised by an effective size ($d$), which is the diameter of a circle with the same area as the projection of all pixels forming the unit; if not stated elsewise, all references to size refer to this effective value. The peak height ($z_{max}$) is the maximum elevation above the target for a given grain. Identification of particles and their sub-units is performed by visual inspection of the calibrated data and, when necessary, cross sections through the 3D data are used, see Extended Data Fig. 4.

For particle E (Fig. 3) a manual levelling of the surface was necessary due to the visible steps (imaging artefacts). Repeating this manual levelling process several times showed that the induced error was negligible. In addition, the height of a grain can only be measured precisely if the grain is directly on the surface and not on another grain. For particle E most of the grains seem to fulfil this requirement, as the mean heights of the grains are smaller than their mean diameters.

### Error analysis

In principle, since AFM tips cannot be infinitely sharp, the size of every particle is overestimated due to the tip sample convolution (i.e. the recorded image reflects a combination of both tip and sample shapes). The convolution uncertainty is generously estimated here to give an upper limit. Since the particle diameter cannot be underestimated by this convolution, the uncertainty interval becomes asymmetric. Values for sizes quoted in the text, and the error bars in Fig. 3 (d) reflect this calculation.

The elongation of particles and grains is calculated by determining its equivalent ellipse (the ellipse with the same second order moments) and choosing the maximum ratio of the largest to smallest of (i) the height of the particle to the major axis, (ii) the height to the minor axis and (iii) the ratio of the major and minor axes. The uncertainties in these ratios take into account the statistical uncertainty due to the manual masking of the particles and the systematic uncertainty due to the tip-sample convolution for the axis lengths. The ratio of the major to minor axis suffers from a large convolution uncertainty which, in some cases (typically particles with steep slopes), prevents a clear statement about their orientation. In these cases no elongation is given. The final uncertainty for the ratio is a worst case estimate which overestimates the uncertainty for non-isolated flat grains.

## Code and data availability

Extended Data Tab. 1 summarises the key parameters for the AFM scans used to produce Fig. 1-3. The filenames listed refer to products available in the ESA Planetary Science Archive where all data used in this paper are freely available. The open source package Gwyddion[28] was used to perform calibration, grain identification and analysis throughout this paper.

# Extended Data legends

**Extended Data Table 1**: **Scan parameters of the primary AFM topography scans discussed in the paper as Figures 1-3**. The number of pixels and thus the pixel resolution at a given scan size was limited by the time available and chosen to maximize the resolution. The filename corresponds to that used in the Planetary Science Archive.

**Extended Data Figure 1: The geometry of the exposures where particle A, B and C were collected.** All exposures are marked by green bars. The top panel shows the distance of Rosetta from the comet (red) and the off-nadir angle (blue). The lower panel shows the latitude and longitude in red and blue. The heliocentric distance during this exposure was 2.25 au.

**Extended Data Figure 2: The geometry of the exposures where particle D was collected.** All exposures are marked by green bars. The top panel shows the distance of Rosetta from the comet (red) and the off-nadir angle (blue). The lower panel shows the latitude and longitude in red and blue. The heliocentric distance during this exposure varied between 2.54 and 2.41 au.

**Extended Data Figure 3: The geometry of the exposures where particle E was collected.** All exposures are marked by green bars. The top panel shows the distance of Rosetta from the comet (red) and the off-nadir angle (blue). The lower panel shows the latitude and longitude in red and blue, respectively. The heliocentric distance during this exposure varied between 2.85 and 2.52 au.

**Extended Data Figure 4: Topographic cross-sections demonstrating the identification of sub-units.** (a) Topographic image of particles A, B and C. Dashed blue, red and green lines show where the cross sections of particle A, B and C, respectively, were made. (b) Height profiles of the 3 cross sections shown in (a) demonstrating how sub-grains were identified, see blue and green arrows, and also revealing the slope of 60 - 70° with the substrate surface.

**Extended Data Figure 5: Tip-sample convolution effects.** Comparison of a spherical particle imaged with an ideal delta-shaped tip (a) and a cone - shaped tip with an opening angle of 30° (b), which is similar to that of the MIDAS tips[14] showing a simulated AFM image (where the colour scale indicates the height) and (c)

an (d) showing the corresponding cross-section through the centre of the structure. The black dashed curve shows the spherical particle, while the blue line depicts the topography as measured with the delta-shaped and pyramidal-shaped tip, respectively. The measurement of the volume of the spherical particle is exaggerated by 25% for the delta-shaped tip and by 50 % for the described cone-shaped tip. The height measurement is not affected by the tip sample convolution.

|  | Figure 1 | Figure 2 | Figure 3 |
| --- | --- | --- | --- |
| target | 14 | 12 | 12 |
| cantilever | 9 | 9 | 7 |
| image resolution | 256 x 256 | 256 x 256 | 192 x 192 |
| image size | 80 x 80 µm² | 20 x 20 µm² | 40 x 40 µm² |
| pixel resolution | 312 nm | 80 nm | 210 nm |
| z step size | 0.7 nm | 0.7 nm | 0.7 nm |
| retraction height | 1095 nm | 977 nm | 734 nm |
| duration | 1 day, 05:05:33 | 08:14:15 | 11:16:30 |
| start time | 2015-04-29T05:21:40Z | 2015-03-13T08:44:38Z | 2015-01-18T20:59:28Z |
| filename | IMG_1509813_1512600_054_ZS | IMG_1507001_1508813_005_ZS | IMG_1501323_1504200_013_ZS |

# Extended Data Table 1

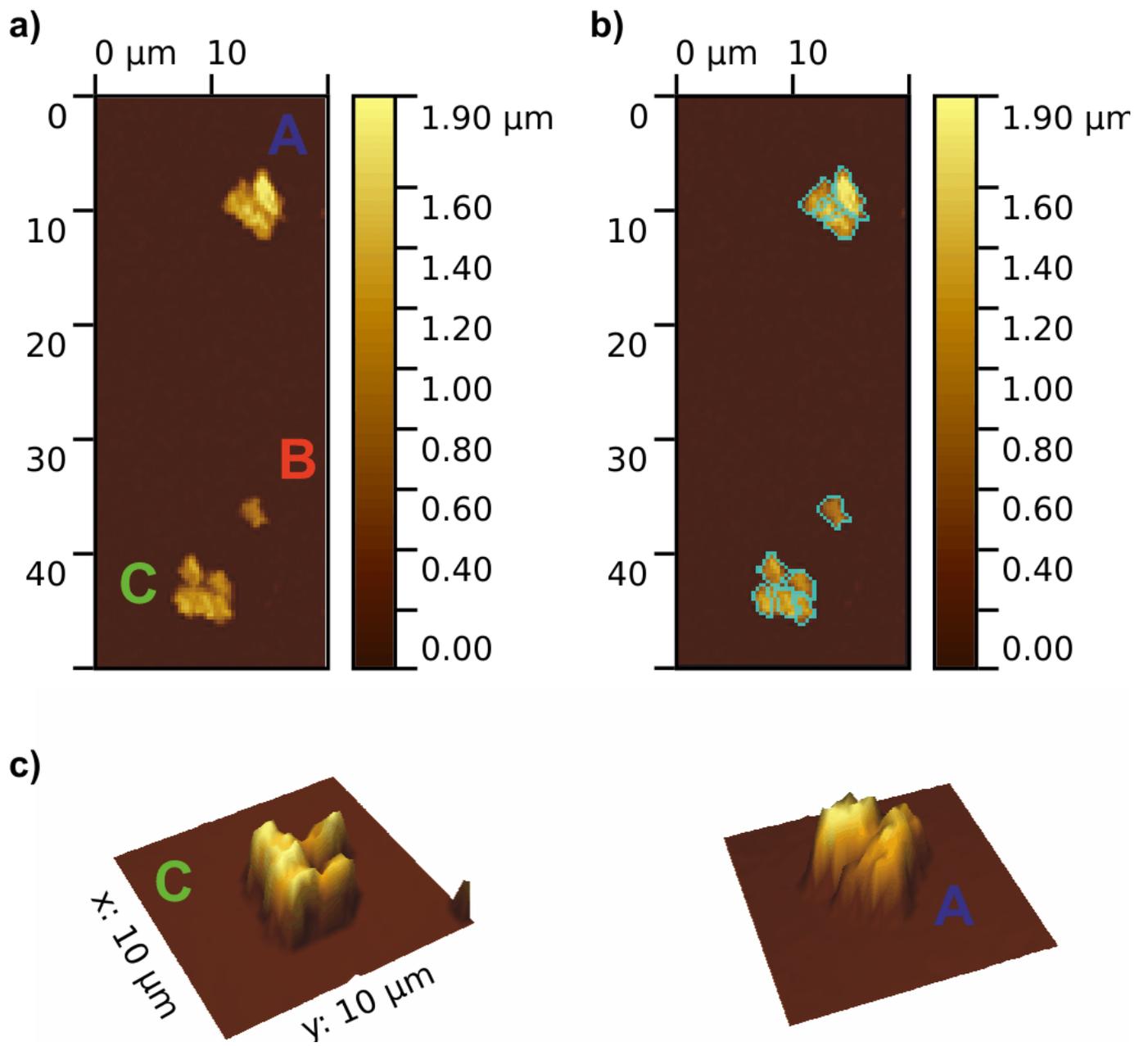

Figure 1

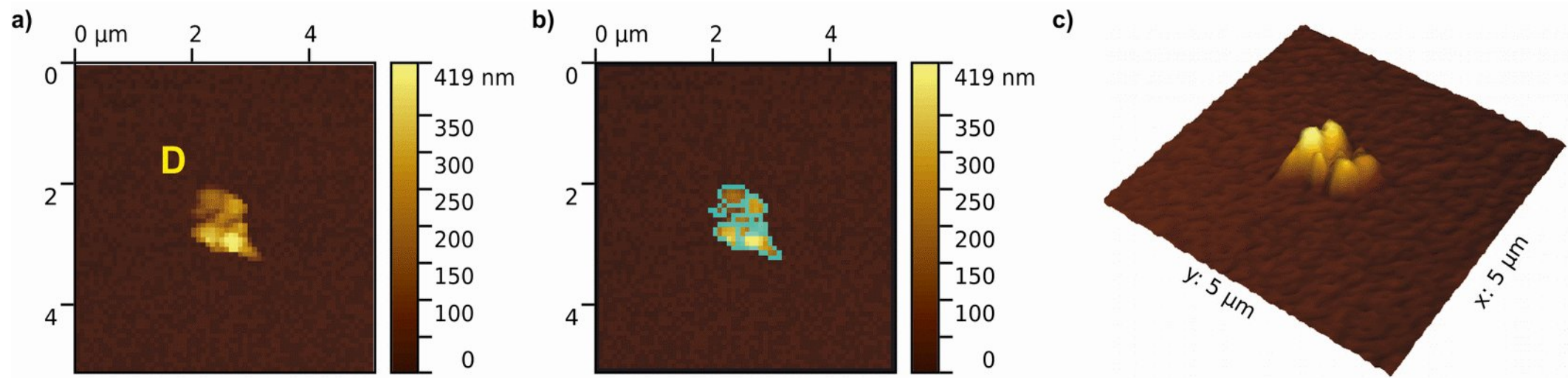

Figure 2

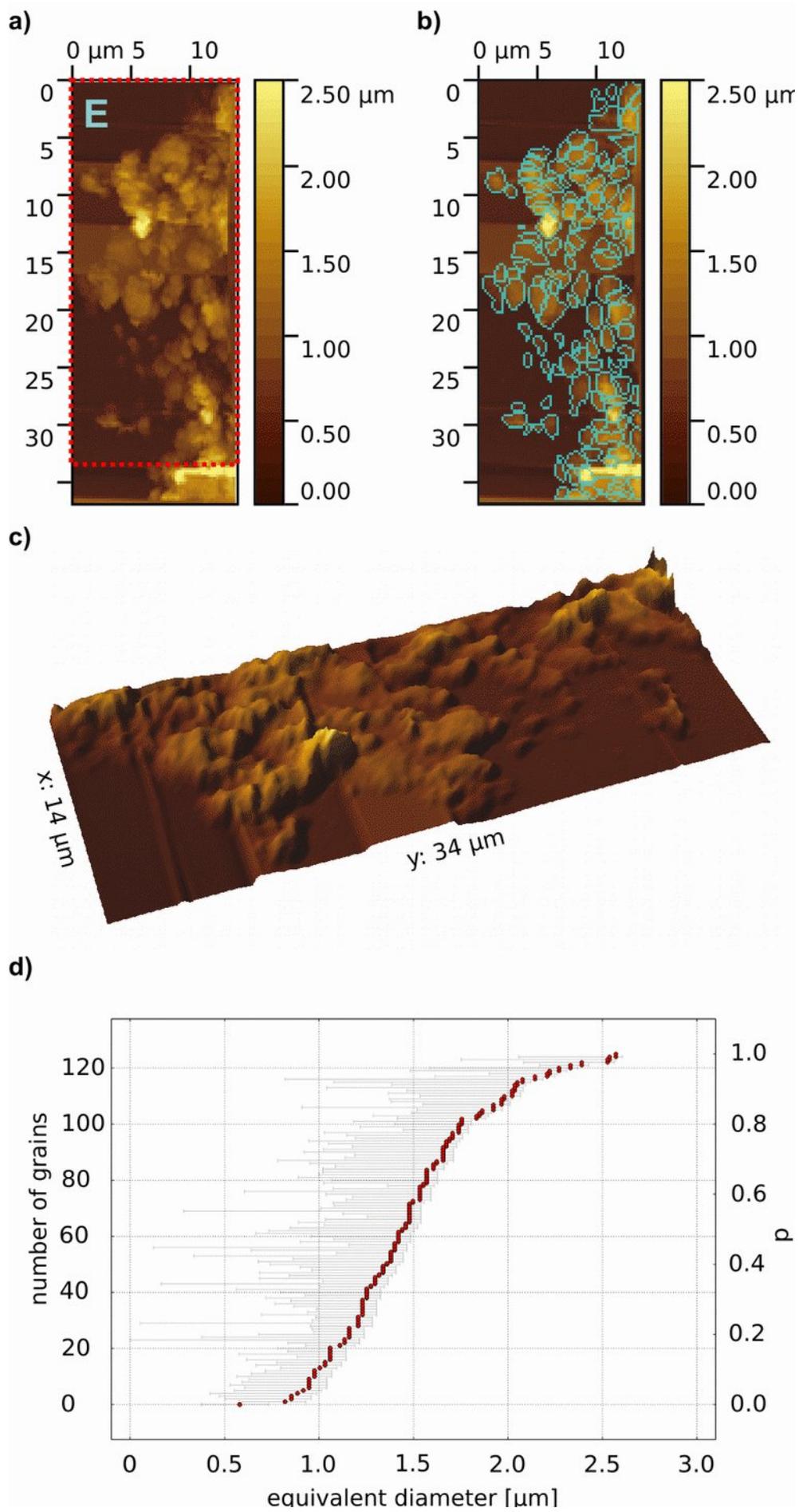

Figure 3

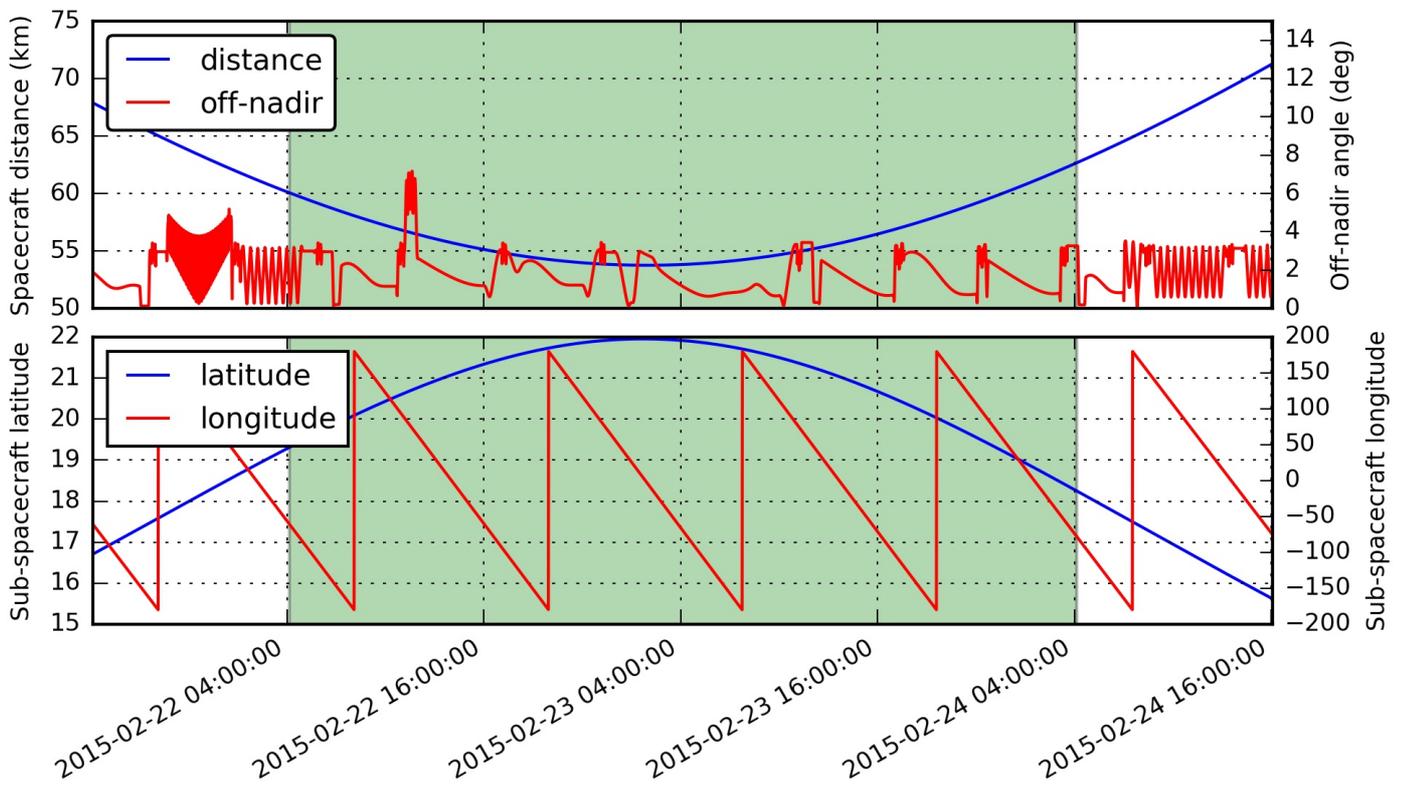

Extended Data Figure 1

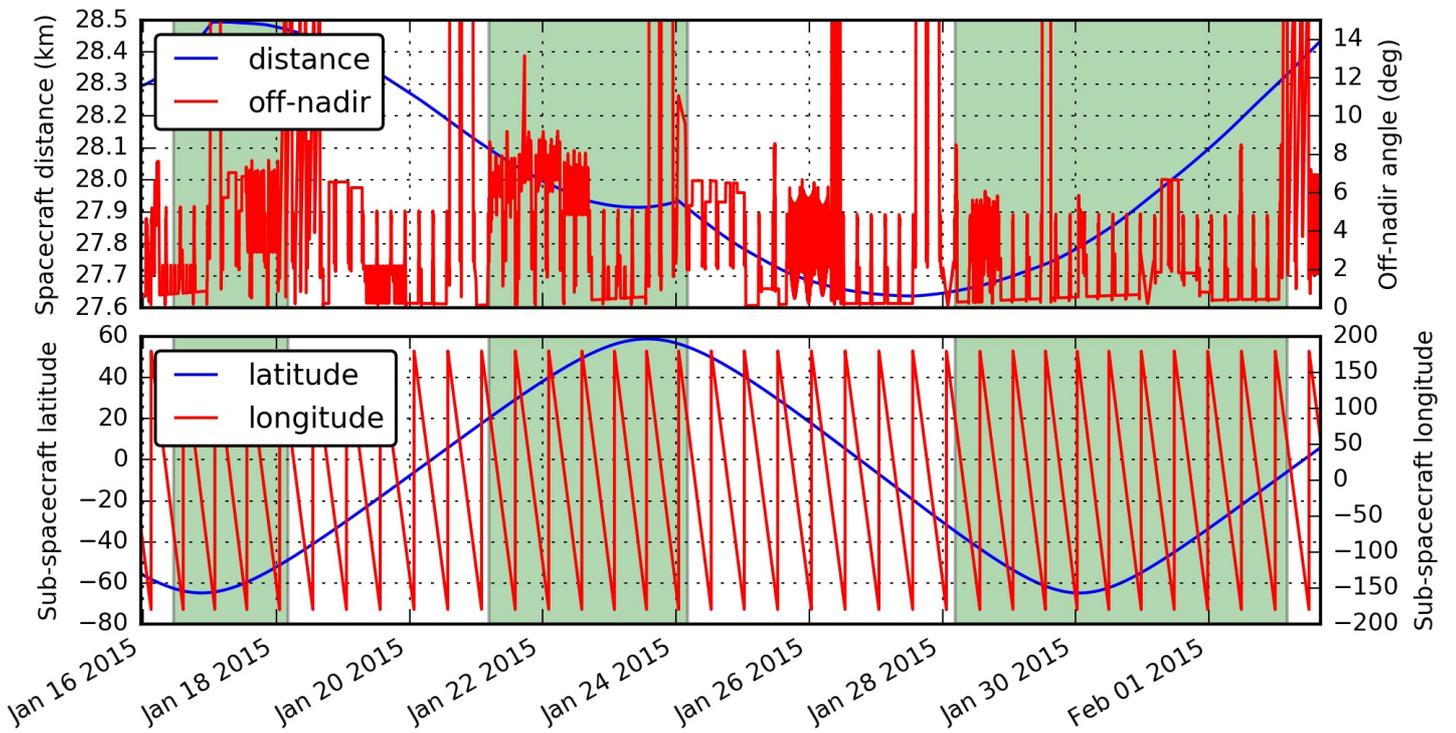

Extended Data Figure 2

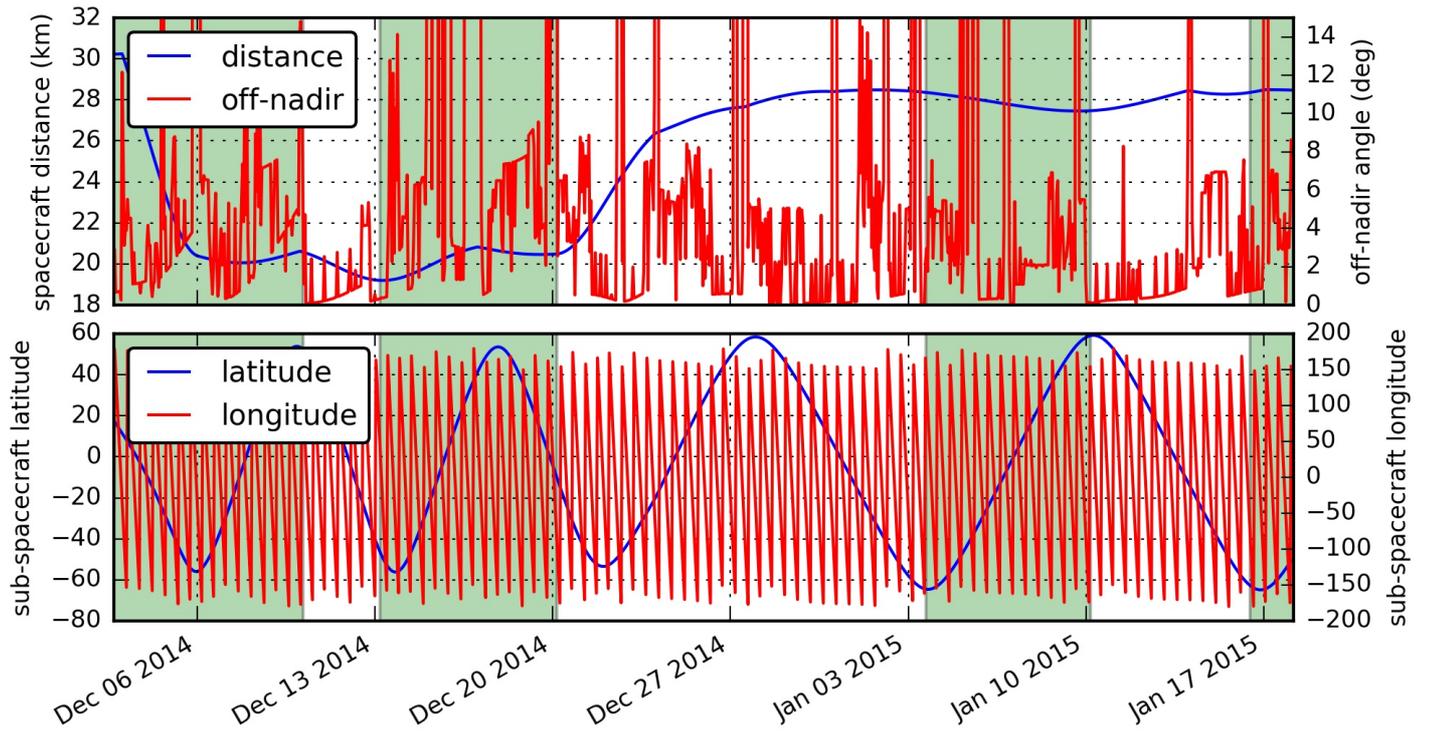

Extended Data Figure 3

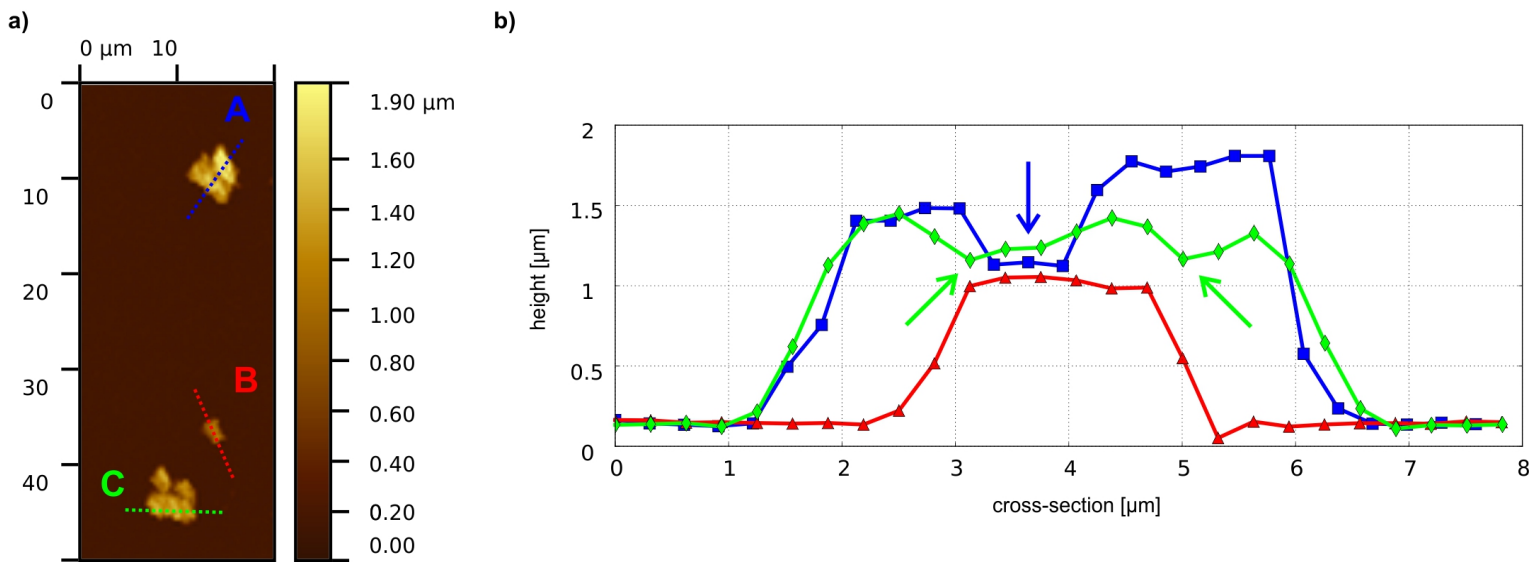

Extended Data Figure 4

a) 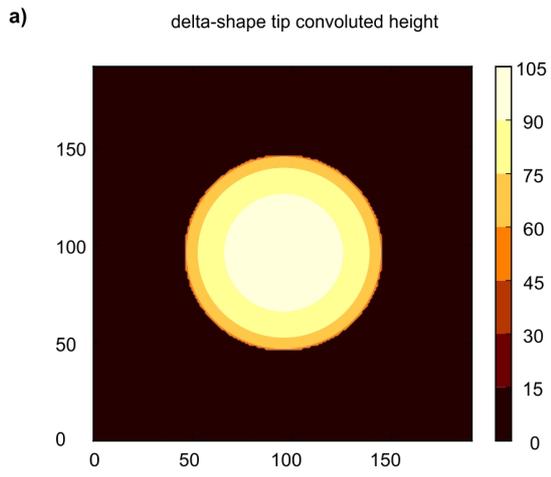

b) 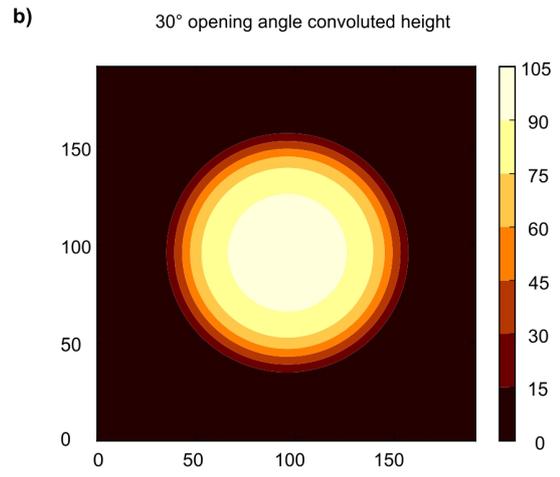

c) 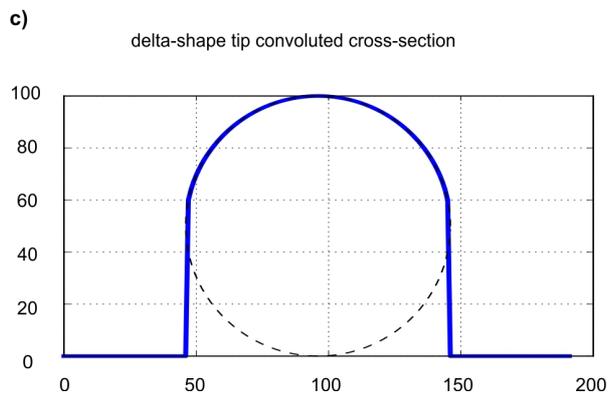

d) 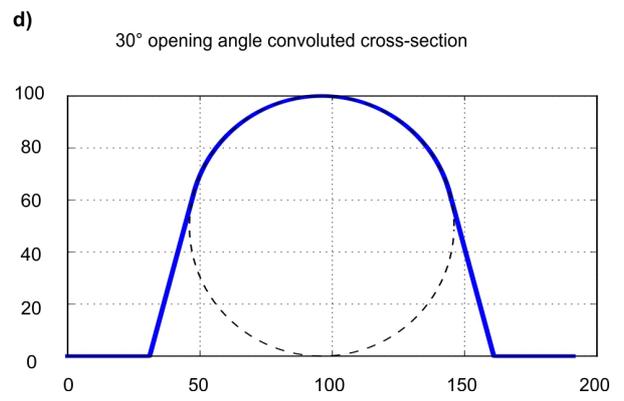

Extended Data Figure 5